\documentclass[12pt,thmsa]{article}
\usepackage{sw20aip}
\input{tcilatex}
\begin{document}

\title{An Adaptive, Kink-Based Approach to Path Integral Calculations}
\author{Randall W. Hall \\
Departments of Chemistry and Physics\\
Louisiana State University\\
Baton Rouge, Louisiana 70803-1804\\
\\
\\
PACS numbers: 31.15.Kb,31.25.-v,02.70.Lq}
\maketitle
\newpage
\begin{abstract}
A kink-based expression for the canonical partition function is
developed using Feynman's path integral formulation of quantum
mechanics and a discrete basis set. The approach is exact for a
complete set of states. The method is tested on the 3x3 Hubbard
model and overcomes the sign problem seen in traditional path
integral studies of fermion systems. Kinks correspond to
transitions between different N-electron states, much in the same
manner as occurs in configuration interaction calculations in
standard \textit{ab initio} methods. The different N-electron
states are updated, based on which states occur frequently during
a Monte Carlo simulation, giving better estimates of the true
eigenstates of the Hamiltonian.
\end{abstract}
\section{Introduction}

Studies of disordered and/or finite size electronic systems, such as clusters, amorphous solids,
or quantum dots pose challenges for computational methods.  The difficulties include the need
to accurately treat electron-electron correlation and to include finite temperature effects (particularly
as they affect the atomic positions.)
The present work is motivated by the desire to simulate large, multi-electron clusters.
Studies of these systems are hindered by
the need to
 identify and/or explore global and local minima. The rapid growth in the number of
minima with cluster size makes the development of precise,
accurate and fast computational algorithms essential for the study
of large systems. Often, isomers have similar energies, which
requires both the calculation of free energies and the accurate
inclusion of electron-electron correlation in order to have a
viable method. Feynman's path integral (PI) formulation of quantum
mechanics has the potential to provide such a method and is the
focus of this work, as its features include exact inclusion of
correlation and the calculation of the partition function, which
allows both the correct sampling of different geometries and the
simultaneous treatment of electronic and geometric degrees of
freedom. The latter is a major advantage, as it does not require
an accurate estimate of the electronic energy at each set of
atomic coordinates. To see this, we can write the partition
function for a system with electronic and geometric degrees of
freedom as
\begin{eqnarray}
Q&=&\int \mathbf{dR}^N \sum_{\{\alpha\}} \rho(\mathbf{R}^N ,
\{\alpha\}, \beta)
\end{eqnarray}
where $\mathbf{R}^N $ denotes the set of geometric coordinates,
$\{\alpha\}$ denotes the electronic basis set, and $\rho$ is the
canonical partition function at an inverse temperature $\beta$. If
this is evaluated by sampling $\{\mathbf{R}^N\}$ and $\{\alpha\}$
from $|\rho|$, we see that new geometries can be sampled without
having a converged electronic state ($\{\alpha\}$ does not need to
correspond to the ground or excited state).  Effectively, the
electronic and geometric optimizations can be carried out
\textit{simultaneously}, in contrast to conventional electronic
optimization methods, in which a converged energy or set of forces
is necessary when changing geometries. The result is that, in
principle, the sampling of geometric phase space can be
accomplished in a time on the order of the time required for a
single electronic energy calculation.

 Most path integral approaches use a complete set
 of states, namely, the position representation, which offers a
 route to an ''exact'' determination of the partition function for
 a many-body system\cite{piMak1,piMak2,piMak3,piOkazaki,ceperleypiH,kukipi,rwhpihub,rwhpiks,ceperley3he,voth1,voth2,ceperleyh,ceperleyfixednode,fandh,AFMC1,AFMC2,AFMC3,AFMC4}. These
approaches have been frustrated by the
 sign of the fermion density matrix, which can be positive or negative and can lead to
 computational difficulties in obtaining precise results. Thus, before the PI  method can
be applied to the study of clusters and other multiple-minima systems, it is necessary
to develop ways to avoid the sign problem.  The sign problem is independent of
whether or not the atomic positions are fixed. Hence, in this first paper on the
subject, we focus on electronic
degrees of freedom with fixed atomic positions and leave the extension to varying
atomic positions to future work.
We note in passing, though, that
many interesting systems have fixed atomic positions (e.g., quantum dots\cite{qdots1}), so that
the developments in this paper can have immediate applications.

Herein, we abandon the use of the position representation and use a
 finite basis set in order to avoid some of the difficulty inherent
with the position representation. It is well known that finite
 basis sets (for example, Gaussian basis sets) are capable of
 producing accurate results for many systems\cite{pople}. We use finite basis
 sets and the discretized version of the path integral expression for Q,
the canonical partition function, to develop a ''kink'' expansion\cite{kink1,kinkwolynes} for Q.
In the discretized version of Q, paths are divided into small imaginary time
segments. When using finite basis sets a path will spend some imaginary
time in one (many-electron) state (this time may involve several imaginary time segments), have a
transition to another state during one imaginary time segment, spend some imaginary time
in the second state, have a transition, etc. The transitions between states are
called ``kinks''\cite{kinkwolynes}. A path can therefore be classified by the number of
kinks and states involved. When we rewrite the expression for Q in
terms of kinks and states, we call this a kink expansion. We note that
kinks can correspond to excitations from the ground (Hartree-Fock)
state
analogous to those seen
 in configuration interaction (CI) calculations. The
expression we develop will give the exact value of Q (including
electron-electron correlation) for a complete set of states. The
zero kink contribution to Q will have no sign problems, since the
system is in a single state. With a properly chosen ground state,
it is possible that convergence of the path
 integral can be obtained with just a few kinks; this would
 significantly reduce the sign problem, thereby increasing
the speed and precision of a calculation.
A good choice
 for the ground state can be obtained using a Monte Carlo
 simulation, in which the different N-electron states that appear
 during the simulation are used to update the estimates of the
 ground and excited states in a way that includes electron-electron
correlation. We call this approach an adaptive approach, since the
Monte Carlo algorithm allows the estimates for the ground and
excited state wavefunctions to evolve according to the statistical
sampling of the different N-electron states.

In the following, we develop a closed form expression for the
 canonical partition function, cast as a kink expansion.
We apply our formulation to a model problem,
the 2-D Hubbard model and demonstrate the efficacy of the
approach.

\section{Kink Formalism}

The kink formalism described here assumes a set (usually finite) of
orthonormal, N-particle states, which we denote by $\left\{ \alpha
_{i}\right\} $. In terms of these states, the partition function can be
written
\begin{equation}
Q=Tr\left\{ e^{-\beta H}\right\} =\sum_{j}\left\langle \alpha _{j}|e^{-\beta
H}|\alpha _{j}\right\rangle
\end{equation}

\noindent We write this as
\begin{eqnarray}
Q &=&\lim_{P\rightarrow \infty }Q\left( P\right)  \nonumber \\
Q\left( P\right)  &=&\sum_{j_{1},j_{2},...,j_{P}}\left\langle
\alpha _{j_{1}}|\exp(-\frac{\beta}{P}H)|\alpha
_{j_{2}}\right\rangle \left\langle \alpha
_{j_{2}}|\exp(-\frac{\beta}{P}H)|\alpha _{j_{3}}\right\rangle
\cdots \nonumber \\
&&\left\langle \alpha _{j_{P}}|\exp(-\frac{\beta}{P}H)|\alpha
_{j_{1}}\right\rangle
 \label{trotter}
\end{eqnarray}

\noindent The introduction of P allows high temperature
approximations for the matrix elements. In this form, the sign
problem can be easily seen. Since each of the matrix elements in
Eqn~\ref{trotter} can be positive, negative, or zero, the sign of
the summand can be negative for some sets of
$\left\{\alpha_{j}\right\}$. Thus, during a Monte Carlo
simulation, the estimator for Q can change sign, leading to large
statistical errors. This alternation of sign is the source of
difficulties in using the path integral formulation to study
multi-fermion systems. We focus on an expression for $Q\left(
P\right) $. The kink expansion is obtained by recasting the sum
over $\left\{ \alpha _{j}\right\} $ as a sum with all $j_{k}$
equal (no kinks), one $j_{k}$ different (2 kinks), two $j_{k}$
different (3 kinks), etc:

\begin{eqnarray}
Q\left( P\right) &=&\sum_{j_{1}}\left[ \left\langle \alpha
_{j_{1}}|\exp(-\frac{\beta}{P}H)|\alpha _{j_{1}}\right\rangle
\right] ^{P}+  \nonumber
\\
&&\sum_{j_{1},j_{2}}\sum_{n=0}^{P-2}\left[ \left\langle \alpha
_{j_{1}}|\exp(-\frac{\beta}{P}H)|\alpha _{j_{1}}\right\rangle
\right] ^{n}\left[ \left\langle \alpha
_{j_{2}}|\exp(-\frac{\beta}{P}H)|\alpha _{j_{2}}\right\rangle
\right] ^{P-2-n}\times  \nonumber \\
&&\left[ \left\langle \alpha
_{j_{1}}|\exp(-\frac{\beta}{P}H)|\alpha _{j_{2}}\right\rangle
\right] ^{2}+\cdots
\end{eqnarray}
where the first term is the zero kink term and the second the two
kink term. In this and the following equations, $j_{1}\neq j_{2}, j_{2}\neq j_{3},$ etc. If we set
\begin{equation}
\epsilon = \beta / P
\end{equation},
\begin{equation}
x_{j}=\left\langle \alpha _{j}|\exp(-\epsilon H)|\alpha
_{j}\right\rangle
\end{equation}

\noindent and
\begin{equation}
t_{ij}=\left\langle \alpha _{i}|\exp(-\epsilon H)|\alpha
_{j}\right\rangle \label{off-diagonal element}
\end{equation}

\noindent we can write the expression for $Q\left( P\right) $:
\begin{equation}
Q\left( P\right) =\sum_{j}x_{j}^{P}+\sum_{n=2}^{P}\left(
\prod_{i=1}^{n}\sum_{j_{i}}\right) \left(
\prod_{k=1}^{n}t_{j_{k},j_{k+1}}\right) \left(
\prod_{k=1}^{n}\sum_{l_{k}=0}^{P-n}\right) \left(
\prod_{k=1}^{n}x_{j_{k}}^{l_{k}}\right) \delta _{l_{1}+l_{2}+\cdots
+l_{n},P-n}
\end{equation}

\noindent where the first term is the zero kink term and $j_{P+1}\equiv
j_{1} $. Recognizing that there are $\binom{P}{n}$ways to put the n kinks at
the different P sites, we can choose the location of the first kink and
rewrite our expression as
\begin{eqnarray}
Q\left( P\right) &=&\sum_{j}x_{j}^{P}+  \nonumber \\
&&\sum_{n=2}^{P}\frac{P}{n}\left(
\prod_{i=1}^{n}\sum_{j_{i}}\right) \left(
\prod_{k=1}^{n}t_{j_{k},j_{k+1}}\right) \times  \nonumber \\
&&\sum_{l_{n}=0}^{P-n}\sum_{l_{n-1}=0}^{P-n-l_{n}}\cdots
\sum_{l_{2}=0}^{P-n-l_{n}-l_{n-1}-\cdots
l_{3}}x_{j_{n}}^{l_{n}}x_{j_{n-1}}^{l_{n-1}}\cdots
x_{j_{2}}^{l_{2}}x_{j_{1}}^{P-n-l_{n}-l_{n-1}-\cdots -l_{2}}
\end{eqnarray}

\noindent With the shorthand notation $S_{j}\equiv l_{n}+l_{n-1}+\cdots
+l_{j}$, we have
\begin{eqnarray}
Q\left( P\right) &=&\sum_{j}x_{j}^{P}+  \nonumber \\
&&\sum_{n=2}^{P}\frac{P}{n}\left(
\prod_{i=1}^{n}\sum_{j_{i}}\right) \left(
\prod_{k=1}^{n}t_{j_{k},j_{k+1}}\right) \times  \nonumber \\
&&\sum_{l_{n}=0}^{P-n}\sum_{l_{n-1}=0}^{P-n-S_{n}}\cdots
\sum_{l_{2}=0}^{P-n-S_{3}}x_{j_{n}}^{l_{n}}x_{j_{n-1}}^{l_{n-1}}\cdots
x_{j_{2}}^{l_{2}}x_{j_{1}}^{P-n-S_{2}}
\end{eqnarray}

Consider the term
\begin{equation}
Q_{n}=\frac{P}{n}\left( \prod_{i=1}^{n}\sum_{j_{i}}\right) \left(
\prod_{k=1}^{n}t_{j_{k},j_{k+1}}\right)
\sum_{l_{n}=0}^{P-n}\sum_{l_{n-1}=0}^{P-n-S_{n}}\cdots
\sum_{l_{2}=0}^{P-n-S_{3}}x_{j_{n}}^{l_{n}}x_{j_{n-1}}^{l_{n-1}}\cdots
x_{j_{2}}^{l_{2}}x_{j_{1}}^{P-n-S_{2}}
\end{equation}

\noindent We first assume that $x_{j_{1}}\neq x_{j_{2}}\neq \cdots \neq
x_{j_{n}}$ and define
\begin{equation}
S\left( \left\{ x_{j}\right\} ,n\right)
=\sum_{l_{n}=0}^{P-n}\cdots
\sum_{l_{2}=0}^{P-n-S_{3}}x_{j_{n}}^{l_{n}}x_{j_{n-1}}^{l_{n-1}}\cdots
x_{j_{2}}^{l_{2}}x_{j_{1}}^{P-n-S_{2}}  \label{Sdefinition}
\end{equation}
\noindent Using,
\begin{equation}
\sum\limits_{l=0}^{M}\QOVERD( ) {x_{i}}{x_{1}}^{l}=\frac{1-\QOVERD( )
{x_{i}}{x_{1}}^{M+1}}{1-\QOVERD( ) {x_{i}}{x_{1}}}
\end{equation}

\noindent we find
\begin{eqnarray}
\sum\limits_{l_{2}=0}^{P-n-S_{3}}x_{j_{2}}^{l_{2}}x_{j_{1}}^{P-n-S_{2}}
&=&\sum\limits_{l_{2}=0}^{P-n-S_{3}}\QOVERD( )
{x_{j_{2}}}{x_{j_{1}}}^{l_{2}}x_{j_{1}}^{P-n-S_{3}}  \nonumber \\
&=&x_{j_{1}}^{P-n-S_{3}}\left( \frac{1-\QOVERD( )
{x_{j_{2}}}{x_{j_{1}}}^{P-n-S_{3}+1}}{1-\QOVERD( ) {x_{j_{2}}}{x_{j_{1}}}}%
\right)  \nonumber \\
&=&\frac{x_{j_{1}}^{P-n-S_{3}+1}-x_{j_{2}}^{P-n-S_{3}+1}}{x_{j_{1}}-x_{j_{2}}%
}  \nonumber \\
&=&\sum_{k=1}^{2}\frac{x_{j_{k}}^{P-n-S_{3}+1}}{\prod\limits_{m\neq
k}^{2}(x_{j_{k}}-x_{j_{m}})}  \nonumber \\
&\equiv &S\left( \left\{ x_{j}\right\} ,2,n\right)
\end{eqnarray}

\noindent We now proceed by induction to develop a general form for $S\left(
\left\{ x_{j}\right\} ,n\right) $. Assume that
\[
S\left( \left\{ x_{j}\right\} ,i-1,n\right) =\sum_{k=1}^{i-1}\frac{%
x_{j_{k}}^{P-n-S_{i}+(i-2)}}{\prod\limits_{k\neq
m}^{i-1}(x_{j_{k}}-x_{j_{m}})}
\]
Consider the next summation in Eq.~\ref{Sdefinition}:
\begin{eqnarray}
&&S\left( \left\{ x_{j}\right\} ,i,n\right)
\sum_{l_{i}=0}^{P-n-S_{i+1}}x_{j_{i}}^{l_{i}}\sum_{k=1}^{i-1}\frac{%
x_{j_{k}}^{P-n-S_{i}+(i-2)}}{\prod\limits_{k\neq
m}^{i-1}(x_{j_{k}}-x_{j_{m}})}  \nonumber \\
&=&\sum_{l_{i}=0}^{P-n-S_{i+1}}x_{j_{i}}^{l_{i}}S\left( \left\{
x_{j}\right\} ,i-1,n\right)  \nonumber \\
&=&\sum_{l_{i}=0}^{P-n-S_{i+1}}x_{j_{i}}^{l_{i}}\sum_{k=1}^{i-1}\frac{%
x_{j_{k}}^{P-n-S_{i+1}-l_{i}+(i-2)}}{\prod\limits_{k\neq
m}^{i-1}(x_{j_{k}}-x_{j_{m}})}  \nonumber \\
&=&\sum_{k=1}^{i-1}\frac{x_{j_{k}}^{P-n-S_{i+1}+(i-2)}}{\prod\limits_{k\neq
m}^{i-1}(x_{j_{k}}-x_{j_{m}})}\times \frac{1-\QOVERD( )
{x_{j_{i}}}{x_{j_{k}}}^{P-n-S_{i+1}+1}}{1-\QOVERD( ) {x_{j_{i}}}{x_{j_{k}}}}
\nonumber \\
&=&\sum_{k=1}^{i-1}\frac{x_{j_{k}}^{P-n-S_{i+1}+(i-1)}}{\prod\limits_{k\neq
m}^{i}(x_{j_{k}}-x_{j_{m}})}-\sum_{k=1}^{i-1}\frac{%
x_{j_{k}}^{i-2}x_{j_{i}}^{P-n-S_{i+1}+1}}{\left( x_{j_{k}}-x_{j_{i}}\right)
\prod\limits_{k\neq m}^{i-1}(x_{j_{k}}-x_{j_{m}})}
\end{eqnarray}

\noindent Now
\begin{eqnarray}
&&-\sum_{k=1}^{i-1}\frac{x_{j_{k}}^{i-2}x_{j_{i}}^{P-n-S_{i+1}+1}}{\left(
x_{j_{k}}-x_{j_{i}}\right) \prod\limits_{m\neq
k}^{i-1}(x_{j_{k}}-x_{j_{m}})}
\nonumber \\
&=&\frac{x_{j_{i}}^{P-n-S_{i+1}+(i-1)}}{\prod\limits_{k\neq
i}^{i}(x_{j_{i}}-x_{j_{k}})}\times \sum_{k=1}^{i-1}\frac{\QOVERD( )
{x_{j_{k}}}{x_{j_{i}}}^{i-2}\prod\limits_{m\neq i}^{i}(x_{j_{i}}-x_{j_{m}})}{%
\left( x_{j_{i}}-x_{j_{k}}\right) \prod\limits_{m\neq
k}^{i-1}(x_{j_{k}}-x_{j_{m}})}  \nonumber \\
&=&\frac{x_{j_{i}}^{P-n-S_{i+1}+(i-1)}}{\prod\limits_{k\neq
i}^{i}(x_{j_{i}}-x_{j_{k}})}\times \sum_{k=1}^{i-1}\frac{\QOVERD( )
{x_{j_{k}}}{x_{j_{i}}}^{i-2}\prod\limits_{m\neq k}^{i-1}(x_{j_{i}}-x_{j_{m}})%
}{\prod\limits_{m\neq k}^{i-1}(x_{j_{k}}-x_{j_{m}})}  \nonumber \\
&=&\frac{x_{j_{i}}^{P-n-S_{i+1}+(i-1)}}{\prod\limits_{k\neq
i}^{i}(x_{j_{i}}-x_{j_{k}})}\times \frac{1}{x_{j_{i}}^{i-2}}\sum_{k=1}^{i-1}%
\frac{\prod\limits_{m\neq k}^{i-1}(x_{j_{i}}-x_{j_{m}})}{\prod\limits_{m\neq
k}^{i-1}(1-\frac{x_{j_{m}}}{x_{j_{k}}})}
\end{eqnarray}

Notice that
\begin{equation}
\sum_{k=1}^{i-1}\frac{\prod\limits_{m\neq k}^{i-1}(x_{j_{i}}-x_{j_{m}})}{%
\prod\limits_{m\neq k}^{i-1}(1-\frac{x_{j_{m}}}{x_{j_{k}}})}\equiv
\sum_{k=1}^{i-1}l_{k}(x_{j_{i}})
\end{equation}

\noindent is an i-2 order polynomial in $x_{j_{i}}$. Consider $%
y(x)=\sum\limits_{k=1}^{i-1}l_{k}(x)$. This is an i-2 order polynomial in x.
Since $l_{k}(x_{j_{l}})=x_{j_{l}}^{i-2}\delta _{k,l}$, $y(x)$ is an i-2
order polynomial that has the value $x_{j_{k}}^{i-2}$ at the i-1 points $%
x_{j_{1}},x_{j_{2}},\ldots ,x_{j_{i-1}}$. Since $y(x)$ is an i-2 order
polynomial, we must have $\sum\limits_{k=1}^{i-1}l_{k}(x)=x^{i-2}$.
Therefore,
\begin{equation}
\frac{1}{x_{j_{i}}^{i-2}}\sum_{k=1}^{i-1}\frac{\prod\limits_{m\neq
k}^{i-1}(x_{j_{i}}-x_{j_{m}})}{\prod\limits_{m\neq k}^{i-1}(1-\frac{x_{j_{m}}%
}{x_{j_{k}}})}=1
\end{equation}

\noindent and
\begin{eqnarray}
&&\sum_{k=1}^{i-1}\frac{x_{j_{k}}^{P-n-S_{i+1}+(i-1)}}{\prod\limits_{k\neq
m}^{i}(x_{j_{k}}-x_{j_{m}})}-\sum_{k=1}^{i-1}\frac{%
x_{j_{k}}^{i-2}x_{j_{i}}^{P-n-S_{i+1}+1}}{\left( x_{j_{k}}-x_{j_{i}}\right)
\prod\limits_{k\neq m}^{i-1}(x_{j_{k}}-x_{j_{m}})}  \nonumber \\
&=&\sum_{k=1}^{i}\frac{x_{j_{k}}^{P-n-S_{i+1}+(i-1)}}{\prod\limits_{k\neq
m}^{i}(x_{j_{k}}-x_{j_{m}})} \\
&=&S\left( \left\{ x_{j}\right\} ,i,n\right)  \nonumber
\end{eqnarray}

\noindent Therefore, by induction, we can show that
\begin{equation}
S\left( \left\{ x_{j}\right\} ,n\right) =\sum_{k=1}^{n}\frac{x_{j_{k}}^{P-1}%
}{\prod\limits_{k\neq m}^{n}(x_{j_{k}}-x_{j_{m}})}
\end{equation}

Next consider the case where some of the $x_{j_{k}}$ are equal. The general
result can be inferred by assuming $x_{0}$ occurs twice in the sum:
\begin{equation}
\cdots
\sum_{k=0}^{P-m}x^{k}_{0}\sum_{l_{1}=0}^{P-m-k}x_{1}^{l_{1}}%
\sum_{l_{2}=0}^{P-m-k-S_{1}}x_{2}^{l_{2}}\cdots
\sum_{l_{n}=0}^{P-m-k-S_{n-1}}x_{n}^{l_{n}}\sum_{j=0}^{P-m-k-S_{n}}x_{0}^{j}%
\sum_{i=0}^{P-m-k-S_{n}-j}x_{i}^{i}\cdots  \label{S'}
\end{equation}

\noindent Now
\begin{equation}
\sum_{l=0}^{M}x_{l}^{l}\sum_{k=0}^{M-l}x_{0}^{k}=\sum_{k=0}^{M}x_{0}^{k}%
\sum_{l=0}^{M-k}x_{l}^{l}
\end{equation}

\noindent so Eq.~\ref{S'} becomes
\begin{equation}
\cdots
\sum_{k=0}^{P-m}x^{k}_{0}\sum_{j=0}^{P-m-k}x_{0}^{j}%
\sum_{l_{1}=0}^{P-m-k-j}x_{1}^{l_{1}}\cdots
\end{equation}

\noindent As a result, if there are $s$ identical $x_{0}$, we will arrive at
\begin{eqnarray}
&&\cdots
\sum_{k_{1}=0}^{P-m}\sum_{k_{2}=0}^{P-m-S_{1}}\sum_{k_{3}=0}^{P-m-S_{2}}%
\cdots \sum_{k_{s}=0}^{P-m-S_{k_{s}-1}}x_{0}^{k_{1}+k_{2}+\cdots
+k_{s}}\sum_{l=0}^{P-m-S_{k_{s}}}x_{l}^{l}\cdots  \label{S double prime} \\
&=&\cdots
\sum_{k_{1}=0}^{P-m}\sum_{k_{2}=0}^{P-m-S_{1}}\sum_{k_{3}=0}^{P-m-S_{2}}%
\cdots
\sum_{k_{s-1}=0}^{P-m-S_{k_{s}-2}}%
\sum_{k_{s}=S_{k_{s}-1}}^{P-m}x_{0}^{k_{s}}\sum_{l=0}^{P-m-k_{s}}x_{l}^{l}%
\cdots  \nonumber \\
&=&\cdots
\sum_{k_{s}=0}^{P-m}x_{0}^{k_{s}}\sum_{k_{1}=0}^{k_{s}}%
\sum_{k_{2}=0}^{k_{s}-S_{1}}\cdots
\sum_{k_{s-1}=0}^{k_{s}-S_{k_{s}-2}}\sum_{l=0}^{P-m-k_{s}}x_{l}^{l}\cdots
\nonumber
\end{eqnarray}

\noindent where we used $\sum_{k=o}^{M-N}\sum_{j=k+N}^{M}=\sum_{j=N}^{M}%
\sum_{k=0}^{j-N}$. Let
\begin{equation}
W\left( s,k_{s}\right) \equiv
\sum_{k_{1}=0}^{k_{s}}\sum_{k_{2}=0}^{k_{s}-S_{1}}\cdots
\sum_{k_{s-1}=0}^{k_{s}-S_{k_{s}-2}}
\end{equation}

\noindent Now we assert that $W\left( s,k_{s}\right)
=\binom{k_{s}+s-1}{s-1}$. When $s=2$, $W\left( s,k_{s}\right)
=k_{s}+1=\binom{k_{s}+s-1}{s-1}$. To show this in general, we use
proof by induction. If $W\left( s-1,k_{s}-k_{1}\right)
=\binom{k_{s}-k_{1}+s-2}{s-2}$, then
\begin{eqnarray}
W\left( s,k_{s}\right) &=&\sum_{k_{1}=0}^{k_{s}}\binom{k_{s}-k_{1}+s-2}{s-2}
\nonumber \\
&=&\sum_{k_{1}=0}^{k_{s}}\frac{\left( k_{s}+s-2-k_{1}\right) !}{\left(
s-2\right) !\left( k_{s}-k_{1}\right) !}  \nonumber \\
&=&\sum_{k_{1}=-k_{s}}^{0}\frac{\left( s-2-k_{1}\right) !}{\left( s-2\right)
!\left( -k_{1}\right) !}  \nonumber \\
&=&\sum_{k_{1}=0}^{k_{s}}\frac{\left( s-2+k_{1}\right) !}{\left( s-2\right)
!\left( k_{1}\right) !}  \nonumber \\
&=&\sum_{k_{1}=0}^{k_{s}}\binom{s-2+k_{1}}{s-2}  \nonumber \\
&=&\binom{s-1+k_{s}}{s-1}
\end{eqnarray}

\noindent where the last identity is taken from Gradshteyn and
Ryzhik\cite {G&R}. Thus, Eqn. \ref{S double prime} becomes
\begin{eqnarray}
&&\cdots \sum_{k_{s}=0}^{P-m}\binom{k_{s}+s-1}{s-1}x_{0}^{k_{s}}%
\sum_{l=0}^{P-m-k_{s}}x_{l}^{l}\cdots  \nonumber \\
&=&\cdots \frac{1}{\left( s-1\right) !}\sum_{k_{s}=0}^{P-m}\frac{d^{s-1}}{%
dx_{0}^{s-1}}x_{0}^{k_{s}+s-1}\sum_{l=0}^{P-m-k_{s}}x_{l}^{l}\cdots
\end{eqnarray}

\noindent Thus, we have for $m$ \textbf{distinct} $x_{j_{k}}$'s, each
occurring $s_{j_{k}}$ times
\begin{equation}
S\left( \left\{ x_{j}\right\} ,n,m,\left\{ s_{j}\right\} \right)
=\prod_{k=1}^{m}\left[ \frac{1}{\left( s_{j_{k}}-1\right) !}\frac{%
d^{s_{j_{k}}-1}}{dx_{j_{k}}^{s_{j_{k}}-1}}x_{j_{k}}^{s_{j_{k}}-1}\right]
\sum_{l=1}^{m}\frac{x_{j_{l}}^{P-n+m-1}}{\prod\limits_{i\neq l}\left(
x_{j_{l}}-x_{j_{i}}\right) }  \label{Final Result for S}
\end{equation}

The derivatives can be evaluated recursively and therefore we have a final
expression for $Q$:

\begin{eqnarray}
Q\left( P\right)  &=&\sum_{j}x_{j}^{P}+  \nonumber \\
&&\sum_{n=2}^{P}\frac{P}{n}\left(
\prod_{i=1}^{n}\sum_{j_{i}}\right) \left(
\prod_{k=1}^{n}t_{j_{k},j_{k+1}}\right) S\left( \left\{
x_{j}\right\} ,n,m,\left\{ s_{j}\right\} \right)
\label{finaleqn}
\end{eqnarray}

This expression can be evaluated using a Monte Carlo algorithm in which
both N-electron states and kinks are sampled. While the number of kinks
can grow to be P, in practical calculations it is plausible that a good
choice of
 states will require only a limited number of kinks. If there are 0 or 2
kinks, it is clear that the Monte Carlo estimator will be positive; for more
than 2 kinks, there will be some negative values for the estimator. However,
as the estimate for the ground and excited states improves, the sign problems will be reduced.

\section{Application: 2-D Hubbard Model}

The formalism was applied to the 2-D Hubbard model\cite{hubbardmodel}. In this model,
 the Hamiltonian is given by
\begin{equation}
\mathcal{H}=at+bU
\end{equation}

\noindent where $t$ is a hopping term that allows hopping between nearest neighbor
sites and $U$ is an repulsive energy term that has non-zero contributions
when two
particles are on the same site. In this work, we chose a 3x3 lattice, $%
a=-1.0$, $b=4.0$, and periodic boundary conditions were
\textit{not used}. We applied the kink formalism to several
occupancies of the lattice: $N_{electron}$ = 3, 4, and 5. The
number of up- and down-spin electrons is given in
Table~\ref{table1}. Temperature was chosen to represent
``typical'' temperatures encountered in studies of clusters:
$\beta \Delta E > 10$ where $\Delta E$ is the difference in energy
between the ground and first excited state. This value was chosen
assuming a 0.1-1 eV gap between ground and excited states (a
typical value in small metallic clusters) and a temperatures on
the order of a hundred Kelvin). The values of $\beta$ are given in
Table~\ref{table1}.

The method was made adaptive in the following manner. An initial
set of N-electron states was chosen by diagonalizing the single
particle Hamiltonian. We then calculated and stored in memory all
the matrix elements
$H_{\alpha^{'},\alpha}=\left<\alpha^{'}|H|\alpha\right>$. These
matrix elements can be constructed from:
\begin{eqnarray}
\text{For}\ \ \ | \alpha>&=&\prod_{i=1}^{N_{electron}}|\alpha_i >\\
\left<\alpha^{'}_i |t|\alpha_i \right>&=&\left\{\begin{array}{cc}
1& \alpha^{'}_i ,\alpha_i \ \text{nearest neighbors}\\
 0\ &\text{otherwise}
 \end{array}\right.
\\
\left<\alpha^{'}_1 ,\alpha^{'}_2 |U|\alpha_1 ,\alpha_2
\right>&=&\delta_{1,2}\delta_{\alpha^{'}_1 ,\alpha_1
}\delta_{\alpha^{'}_2 ,\alpha_2 }
\end{eqnarray}

\noindent These were then used to approximate the density matrix
elements as:
\begin{eqnarray}
<\alpha^{'}|\exp(-\epsilon
H)|\alpha>&\approx&<\alpha^{'}|(1-\epsilon H)|\alpha>
\end{eqnarray}
\noindent P was chosen large enough so that the results were
converged to the exact result.
 For larger basis sets, the storage of matrix elements will not
be feasible, but can be calculated ``on-the-fly''. During the
course of the Monte Carlo calculation, we kept track of the
N-electron states that appeared at each step. For the first 5000
steps, we periodically performed diagonalizations using only the
states that appeared at least 0.1\% of the time. Thus, a subset of
the total number of N-electron states were diagonalized, using the
current set of matrix elements $H_{\alpha^{'},\alpha}$. After each
diagonalization, all the Hamiltonian matrix elements were updated.
Therefore, after a number of diagonalizations, the states we
labeled as $\alpha$ were linear combinations of the original set
of N-electron states.  During subsequent Monte Carlo steps, we
expected the adapted states to produce
  fewer kinks and to
reduce the sign problem.

This adaptive procedure is similar to stochastic diagonalization
methods\cite{stochasticdiag}, used to find the lowest eigenvalues of large Hamiltonians.
The major difference
is that here we include states that couple to excited states in the set of states
used for diagonalization (which we must, since we are not focusing solely on obtaining
ground state energies, but rather the partition function). Stochastic diagonalization
corresponds, roughly, to using the present approach, but limiting the number of kinks to 2
(one of the two states restricted to being the ground state)
and performing a diagonalization every time a second kink is accepted.
In addition, since the  overall goal is to
simultaneously sample atomic positions and treat the electronic problem, the Hamiltonian
matrix elements we are concerned with will change during the simulation, which is not the
case in standard stochastic diagonalization techniques.  In an application of the path
integral approach to systems other than model systems, it is likely that many of the
ideas of stochastic diagonalization can be ``borrowed'' for use in the adaptive scheme,
while at the same time utilizing the power of the path integral method to treat
finite temperatures and changing atomic positions.
In any event, the adaptive procedure resulted in a new set of correlated N-electron
states that were better representations of the eigenstates of the Hamiltonian.
After diagonalization, the system was put
in the ground state and the number of kinks was set to 0. After 5000 steps,
15000 additional steps with no further diagonalization
were taken in order to calculate the total energy.

The results are shown in Table~\ref{table1}.  We first rewrote  Eqn.~\ref{finaleqn} as
\begin{eqnarray}
Q(P)&=&\sum^{P}_{n=1}\left(\prod^{n}_{i=1}\sum_{j_{i}}\right)\rho_{n}(\{\alpha_{j}\})
\end{eqnarray}
Here, $n=1$ corresponds to the zero kink term and $\rho$ is the density matrix corresponding
to $n$ and a specific set of states $\{\alpha\}$ We evaluated the average sign:
\begin{eqnarray}
\left<S\right>&=&\frac{\left<\rho_{n}\right>}{\left<|\rho_{n}|\right>}
\end{eqnarray}
For a poorly chosen set of states, $\left<S\right>$ will be nearly zero\cite{kukipi}.
Thus, $\left<S\right>$ is a good measure of the severity of the sign problem.  We also
evaluated the average energy via
\begin{eqnarray}
\left<E_{Monte Carlo}\right>&=& \frac{\left< -\partial \rho_{n} / \partial \beta\right>}
{\left<\rho_{n}\right>}
\end{eqnarray}
by taking the appropriate derivative of Eqn.~\ref{finaleqn}.
An examination of Table~\ref{table1} shows that the average sign of the density
matrix is unity and demonstrates that the adaptive approach does indeed overcome the
sign problem.  An analysis of the number of types of kinks encountered
indicated that only 0 and 2 kinks were present after the 5000 step adaptive
period.  Since the density matrix is always positive for 0 or 2 kinks, this lead to
the value of 1.0 for the average sign.  Without the adaptive procedure, the average
sign would have been very small and the usual sign problems would have
been encountered.

It is interesting to examine the convergence of the ground state energy
to its final value during the adaptive phase of the simulation (the first
5000 Monte Carlo passes), as well as the number of states involved in the
diagonalization. This information is given in Table~\ref{table2} and Table~\ref{table3}.
It is evident that only a subset of the total number of states need to be included at any one diagonalization,
which aids in the speed of the calculation. In addition, the energy is seen to converge
relatively quickly to a final value during the adaptive stage. $N_{electron}$ = 3 and 5
 doubly degenerate ground states, leading to 2 states being ``visited'' a substantial
amount of time. Also note that, after the final diagonalization, the ground
state energy of $N_{electron}$ = 4 is not the exact ground state energy, but
the exact average energy is obtained during the simulation with the addition of
kinks. Finally, we note that the last diagonalization for $N_{electron}$ = 5, makes
a very small change in the total energy.

It may be noticed that, for each value of $N_{electron}$, a large
fraction of the states are involved in one of the diagonalization
steps. This leads to the question of whether this is a required
feature of the adaptive scheme. To test this, we performed an
additional simulation for $N_{electron}$ = 5 and restricted the
number of states that could be diagonalized at any time to 300.
The results are shown in Table~\ref{table4}, from which it is
evident that a diagonalization step involving a large fraction of
the states is not needed for the adaptive scheme to work, although
diagonalizing fewer states at each step does increase the length
of the adaptive period.

\section{Conclusions}
The adaptive approach has been shown to be successful in
overcoming the sign problem inherent in standard path integral
approaches. The use of the closed form expression for the kink
expansion allows large values of P to be used. This method needs
to be applied to additional systems, to assess its overall
utility. In particular, there needs to be an assessment of how the
method scales with the number of basis functions.  This is not
straightforward to determine, since this will depend on what types
of schemes are used to speed up the calculation.  However, it is
likely that, for fixed atomic positions, that the method will
scale in time similar to CI calculations,  though it will not have
to exhaustively evaluate all excitations (singles, doubles, etc.)
and may therefore require  less time than a traditional CI
calculation. In addition, the size of the space of Slater
determinants may grow too large for brute force diagonalization,
while the adaptive approach can still be used (much in the same
way that stochastic diagonalization can be used). Since the
electronic calculation can be carried out simultaneously with
atomic position optimization, the time for a calculation with
varying atomic positions should be comparable to the time required
to perform a \textit{single point} \textit{ab initio} CI
calculation. Additionally, as the adaptive wavefunctions become
better representations of the exact wavefunction, the calculation
should speed up, since only a small number of states will be
coupled to the ground state. Future work will assess the
applicability of this method to such types of calculations.

\section{Acknowledgments}
Calculations were performed on computers purchased with a grant from the Louisiana Education Quality
Support Fund.

\bibliographystyle{hphysrev3}
\bibliography{kink}

\begin{thebibliography}{10}

\bibitem{piMak1}
C.~Mak, R.~Egger, and H.~Weber-Gottschick,
\newblock Phys Rev Lett {\bf 81}, 4533 (1998).

\bibitem{piMak2}
C.~Mak and R.~Egger,
\newblock J Chem Phys {\bf 110}, 12 (1999).

\bibitem{piMak3}
R.~Egger, L.~Muhlbacher, and C.~Mak,
\newblock Phys Rev E {\bf 61}, 5961 (2000).

\bibitem{piOkazaki}
S.~Miura and S.~Okazaki,
\newblock J Chem Phys {\bf 112}, 10116 (2000).

\bibitem{ceperleypiH}
B.~Militzer, W.~Magro, and D.~Ceperley,
\newblock Contributions to plasma physics {\bf 39}, 151 (1999).

\bibitem{kukipi}
W.~Newman and A.~Kuki,
\newblock J Chem Phys {\bf 96}, 1409 (1992).

\bibitem{rwhpihub}
R.~Hall,
\newblock J Chem Phys {\bf 94}, 1312 (1991).

\bibitem{rwhpiks}
R.~Hall and M.~Prince,
\newblock J Chem Phys {\bf 95}, 5999 (1991).

\bibitem{ceperley3he}
D.~Ceperley,
\newblock Phys Rev Lett {\bf 69}, 331 (1992).

\bibitem{voth1}
P.~Roy, S.~Jang, and G.~Voth,
\newblock J Chem Phys {\bf 111}, 5303 (1999).

\bibitem{voth2}
P.~Roy and G.~Voth,
\newblock J Chem Phys {\bf 110}, 3647 (1999).

\bibitem{ceperleyh}
B.~Militzer, W.~Magro, and D.~Ceperley,
\newblock Contrib Plasma Phys {\bf 89}, 151 (1999).

\bibitem{ceperleyfixednode}
D.~Ceperley,
\newblock Path integral monte carlo methods for fermions,
\newblock in {\em Monte Carlo and molecular dynamics of condensed matter
  systems}, edited by K.~Binder and G.~Ciccotti, 1996.

\bibitem{fandh}
R.~P. Feynman and A.~Hibbs,
\newblock {\em Quantum Mechanics and Path Integrals} (McGraw-Hill, 1965).

\bibitem{AFMC1}
N.~Rom, E.~Fattal, A.~K. Gupta, E.~A. Carter, and D.~Neuhauser,
\newblock J. Chem. Phys. {\bf 109}, 8241 (1998).

\bibitem{AFMC2}
N.~Rom, D.~M. Charutz, and D.~Neuhauser,
\newblock Chem. Phys. Lett. {\bf 270}, 382 (1997).

\bibitem{AFMC3}
Y.~Asai,
\newblock Phys. Rev. B {\bf 62}, 10674 (2000).

\bibitem{AFMC4}
R.~Baer, M.~Head-Gordon, and D.~Neuhauser,
\newblock J. Chem. Phys. {\bf 109}, 6219 (1998).

\bibitem{qdots1}
J.~Harting, O.~M{\"{u}}ller, and P.~Borrmann,
\newblock Phys. Rev. B {\bf 62}, 10207 (2000).

\bibitem{pople}
W.~Hehre, L.~Radom, P.~Schleyer, and J.~Pople,
\newblock {\em ab initio molecular orbital theory} (Wiley, 1986).

\bibitem{kink1}
K.~Schotte and V.~Schotte,
\newblock Phys Rev B {\bf 4}, 2228 (1971).

\bibitem{kinkwolynes}
R.~Chiles, G.~Jongeward, M.~Bolton, and P.~Wolynes,
\newblock J Chem Phys {\bf 81}, 2039 (1986).

\bibitem{G&R}
I.~Gradshteyn, I.~Ryzhik, A.~Jeffrey, and D.~Zwillinger,
\newblock {\em Table of Integrals, Series, and Products} (Academic Press,
  2000).

\bibitem{hubbardmodel}
M.~Imada, A.~Fujimori, and Y.~Tokura,
\newblock Rev. Mod. Phys. {\bf 70}, 1039 (1998).

\bibitem{stochasticdiag}
H.~De~Raedt and M.~Frick,
\newblock Physics Reports {\bf 231}, 107 (1993).

\end{thebibliography}
\newpage
\begin{center}Table Captions\end{center}

\noindent Table 1 Caption. Parameters for 3x3 Hubbard Model. $N_{electron}$ is the
total number of electron, $N_{up}$ and $N_{down}$ the number of up- and down-spin
electrons, respectively, $\beta$ is the inverse temperature, $<E_{exact}>$
is the exact thermally averaged energy, $\Delta E$ is the difference between the ground
and first excited states, $P$ is the number of discretization points, $<E_{Monte Carlo}>$
is the average energy from the simulation, and $<S>$ is the average sign of the
density matrix. Numbers in parenthesis represent 2 standard deviations.

\noindent Table 2 Caption. Number of states included in the diagonalization for
the adaptive steps performed during the first 5000 Monte Carlo
passes. Diagonalizations were
performed after every 500 Monte Carlo passes up to 5000 passes. In addition, diagonalizations were performed
after 1 and 100 Monte Carlo steps. The total number of states are 324 (N=3), 1296 (N=4), and 3024 (N=5).

\noindent Table 3 Caption. Convergence of the ground state energy with adaptive diagonalizations.

\noindent Table 4 Caption. Convergence of the ground state energy with adaptive
diagonalization for $N_{electron}$ = 5 and with the maximum number
of states allowed to be diagonalized restricted to 300.
After 5000 passes, diagonalizations were only
performed when more than 2 kinks occurred for a significant amount
of time. Following the adaptive period, the average energy was
calculated and found to be -8.214186 (6).

\newpage
\begin{table}[tbp] \centering%
\begin{tabular}{|c|c|c|c|}
\hline
 &$N_{electron}$ = 3 &$N_{electron}$ = 4 &$N_{electron}$ = 5 \\ \hline
$N_{up}$&2&2&3 \\ \hline
$N_{down}$&1&2&2 \\ \hline
$\beta $& 50.0&200.0&50.0 \\ \hline
$<E_{exact}>$ & -6.50920068& -7.649675&-8.21418803 \\ \hline
$\beta \Delta E$&42.6&46.2&36.0 \\ \hline
$P$& 524288&2097152& 524288 \\ \hline
$<E_{Monte Carlo}>$& -6.50920068(8)& -7.649676(5)& -8.21418801(4) \\ \hline
$<S>$&1.00(0)&1.00(0)&1.00(0) \\ \hline
\end{tabular}
\caption{\label{table1}}
\end{table}
Table \ref{table1} Parameters for 3x3 Hubbard Model.
$N_{electron}$ is the total number of electrons, $N_{up}$ and
$N_{down}$ the number of up- and down-spin electrons,
respectively, $\beta$ is the inverse temperature, $<E_{exact}>$ is
the exact thermally averaged energy, $\Delta E$ is the difference
between the ground and first excited states, $P$ is the number of
discretization points, $<E_{Monte Carlo}>$ is the average energy
from the simulation, and $<S>$ is the average sign of the density
matrix. Numbers in parenthesis represent 2 standard deviations.
\newpage
\begin{table}[tbp]
\begin{tabular}{|c|c|c|c|c|c|c|c|c|c|c|c|c|}
\hline
Pass Number&$N_{electron}$=3&$N_{electron}$=4&$N_{electron}$=5 \\ \hline
   1&  1&  1&   1 \\ \hline
 100&125& 67& 310 \\ \hline
 500&132&453&1404 \\ \hline
1000&  2& 29&1303 \\ \hline
1500&  2&  9&   2 \\ \hline
2000&  2&  5&   2 \\ \hline
2500&  2&  4&   2 \\ \hline
3000&  2&  4&   2 \\ \hline
3500&  2&  3&   2 \\ \hline
4000&  2&  4&   2 \\ \hline
4500&  2&  2&   2 \\ \hline
5000&  2&  4& 604 \\ \hline
\end{tabular}
\caption{\label{table2}}
\end{table}
Table \ref{table2}. Number of states included in the
diagonalization for the adaptive steps performed during the first
5000 Monte Carlo passes. Diagonalizations were performed after
every 500 Monte Carlo passes up to 5000 passes. In addition,
diagonalizations were performed after 1 and 100 Monte Carlo steps.
The total number of states are 324 (N=3), 1296 (N=4), and 3024
(N=5).
\newpage
\begin{table}[tbp]
\begin{tabular}{|c|c|c|c|}
\hline
Pass Number&$N_{electron}$=3&$N_{electron}$=4&$N_{electron}$=5 \\ \hline
   1&-6.13356781&-6.92278137&-7.21199493  \\ \hline
 100&-6.50883238&-7.55718596&-8.14959112  \\ \hline
 500&-6.50920068&-7.64950387&-8.21394605  \\ \hline
1000&-6.50920068&-7.64958381&-8.21418741  \\ \hline
1500&-6.50920068&-7.64962108&-8.21418741  \\ \hline
2000&-6.50920068&-7.64963715&-8.21418741  \\ \hline
2500&-6.50920068&-7.64964181&-8.21418741  \\ \hline
3000&-6.50920068&-7.64964634&-8.21418741  \\ \hline
3500&-6.50920068&-7.64964922&-8.21418741  \\ \hline
4000&-6.50920068&-7.64965215&-8.21418741  \\ \hline
4500&-6.50920068&-7.64965246&-8.21418741  \\ \hline
5000&-6.50920068&-7.64965656&-8.21418802  \\ \hline
\end{tabular}
\caption{ \label{table3}}
\end{table}
Table \ref{table3}. Convergence of the ground state energy with
adaptive diagonalizations.
\newpage
\begin{table}[tbp]
\begin{tabular}{|c|c|c|}
\hline
Pass Number&States Involved&Energy \\ \hline
     1&  1&-7.21199493 \\ \hline
   100&178&-8.12195973\\ \hline
   500&295&-8.17901325\\ \hline
  1000&257&-8.18986053\\ \hline
  1500&274&-8.19522379\\ \hline
  2000&275&-8.19876874\\ \hline
  2500&227&-8.20054427\\ \hline
  3000&283&-8.20333449\\ \hline
  3500&283&-8.20517670\\ \hline
  4000&245&-8.20888475\\ \hline
  4500&236&-8.21053144\\ \hline
  5000&213&-8.21132950\\ \hline
  5500&294&-8.21260450\\ \hline
  6000&300&-8.21349827\\ \hline
  6500&293&-8.21396900\\ \hline
  7000&289&-8.21407486\\ \hline
  7500&  6&-8.21407824\\ \hline
  8000&  3&-8.21407828\\ \hline
  8500&284&-8.21412725\\ \hline
  9000&202&-8.21414318\\ \hline
  9500&  3&-8.21414335\\ \hline
 10000&  4&-8.21414377\\ \hline
 12000&221&-8.21415918\\ \hline
 12500&231&-8.21417133\\ \hline
 14500&199&-8.21417807\\ \hline
\end{tabular}
\caption{\label{table4}}
\end{table}
Table \ref{table4}. Convergence of the ground state energy with
adaptive diagonalization for $N_{electron}$ = 5 and with the
maximum number of states allowed to be diagonalized restricted to
300. After 5000 passes, diagonalizations were only performed when
more than 2 kinks occurred for a significant amount of time.
Following the adaptive period, the average energy was calculated
and found to be -8.214186 (6).
\end{document}